\documentclass[12pt]{article}
\usepackage{epsfig,amssymb,cite,multirow} 
\usepackage{amsmath}

\usepackage{amscd}
\usepackage{amsmath,graphicx}
\usepackage{xypic}
\usepackage[matrix,arrow,curve]{xy}
\usepackage{graphicx}
\usepackage{verbatim}
\usepackage{epsf}
\usepackage{latexsym}
\usepackage{amsmath,amsfonts,amssymb,amsthm}
\usepackage{amsmath,amsthm}

\def\bseq{\begin{subequation}}  % = 1a 1b
\def\eseq{\end{subequation}}
\def\bsea{\begin{subeqnarray}}  % = 1.1a 1.1b
\def\esea{\end{subeqnarray}}
\def\Tilde#1{\widetilde{#1}}                    % big tilde

\newcommand{\bbox}{\lower.2ex\hbox{$\Box$}}

\newcommand{\beq}{\begin{equation}}
\newcommand{\eeq}{\end{equation}}
\newcommand{\bea}{\begin{eqnarray}}
\newcommand{\eea}{\end{eqnarray}}
\newcommand{\ena}{\end{eqnarray}}

\newcommand{\D}{\Delta}

\newcommand{\Tr}{{\rm Tr}}

\newcommand{\be}{\begin{equation}}
\newcommand{\ee}{\end{equation}}

\def\D{{\cal D}}

\begin{document}
\setcounter{page}{0}
\begin{titlepage}
\titlepage
\begin{flushright}
LPTENS-09/35\\
\end{flushright}
\begin{center}
\LARGE{\Huge ${\cal N}=1$ Chern-Simons  theories, orientifolds and Spin(7) cones\\}
\LARGE{\Huge  }
\end{center}
\vskip 1.5cm \centerline{{\bf Davide Forcella$^{a}$ \footnote{\tt forcella@lpt.ens.fr}
  and Alberto Zaffaroni$^{b}$ \footnote{\tt alberto.zaffaroni@mib.infn.it}
%..................$^{c}$ \footnote{\tt ........................}
}}
%\medskip
\vskip 1cm
\footnotesize{

\begin{center}
$^a$ Laboratoire de Physique Th\'eorique de l'\'Ecole Normale Sup\'erieure \\
and CNRS UMR 8549\\
24 Rue Lhomond, Paris 75005, France\\
%\medskip
%$^b$ PH-TH Division, CERN 
%CH-1211 Geneva 23, Switzerland\\
\medskip
$^b$  Universit\`a di Milano Bicocca and INFN, sezione di
Milano-Bicocca \\
piazza della Scienza 3, Milano 20126, Italy
\end{center}}

\bigskip

\begin{abstract}

We construct three dimensional ${\cal N}=1$ Chern-Simons theories living on M2 branes probing Spin(7) cones.
We consider  Spin(7) manifolds obtained as quotients of Calabi-Yau four-folds by an anti-holomorphic involution, following a construction by Joyce. The corresponding Chern-Simons theories can be obtained from ${\cal N}=2$
theories by an orientifolding procedure. These theories are holographically dual to M theory solutions $AdS_4\times H$, where the weak $G_2$  manifold $H$ is the base of the Spin(7) cone.

\end{abstract}

\vfill
\begin{flushleft}
{\today}\\
\end{flushleft}
\end{titlepage}

\newpage

\tableofcontents

\section{Introduction}

There has been some progress in understanding the conformal field theory living on a stack of N M2 branes at the tip of a non compact  eight-dimensional cone. This theory is the holographic dual of a Freund-Rubin solution in M theory of the form $AdS_4\times H$, where $H$ is the seven-dimensional  compact base of the cone.  In the case of large
supersymmetry ${\cal N}\ge 3$ the conformal field theory has been identified with a Chern-Simons theory \cite{Aharony:2008ug, Benna:2008zy, Hosomichi:2008jb, Jafferis:2008qz, Aharony:2008gk}.  The  case where the cone is a Calabi-Yau four-fold corresponds to  ${\cal N}=2$ 
supersymmetry. A general construction of ${\cal N}=2$ 
Chern-Simons theories dual to Calabi-Yau cones  has been discussed in \cite{Martelli:2008si,Hanany:2008cd} and 
a large number of models have been subsequently constructed  \cite{Hanany:2008fj,Ueda:2008hx, Imamura:2008qs, Franco:2008um, Hanany:2008gx, Davey:2009sr, Franco:2009sp,Amariti:2009rb, Martelli:2009ga }. Much less is known about ${\cal N}=1$ theories and some recent studies can be found in \cite{Mauri:2008ai,Ooguri:2008dk,Gaiotto:2009mv,Bobev:2009ms}. Actually three dimensional $\mathcal{N}=1$ superconformal field theories are particularly interesting because they are still supersymmetric but they do not have any holomorphic property. This is a peculiarity of the AdS$_4$/CFT$_3$ correspondence with respect to the usual AdS$_5$/CFT$_4$, and it is worthwhile to study it.
   
In this paper we are interested in  the three dimensional conformal field theory living on N M2 branes at the tip of Spin(7) cones. This case corresponds to  ${\cal N}=1$ supersymmetry.  We will use a general  construction due to Joyce \cite{Joyce:1999nk} and we will realize Spin(7) cones $X$  as  quotients of  four dimensional Calabi Yau  cones $Y$ by an  anti-holomorphic involution $\Theta$: $X=Y/\Theta$.  We will obtain theories corresponding to  Spin(7) cones $X=Y/\Theta$ by performing an orientifold  quotient of the ${\cal N}=2$ Chern-Simons theory corresponding to $Y$. The resulting theory will have a residual ${\cal N}=1$ supersymmetry. We will show that the moduli space for one membrane
is  naturally a quotient of $Y$ by a dihedral group $\Theta_k$, which is generated by the anti-involution and an abelian symmetry $\mathbb{Z}_{2k}$. This is a quite general construction of Chern-Simons theory with Spin(7) moduli space.  The case of anti-involutions without fixed points, except the tip of the cone, is the most interesting one; in this case
the near horizon background $AdS_4\times H/\Theta_k$ is perfectly smooth. We will provide many examples
of ${\cal N}=1$ Chern-Simons corresponding to anti-involutions without fixed points on the base.  

The orbifolding and orientifolding procedure are standard for theories obtained from D branes in
type II string theories but are less understood for M2 branes in M theory. From the string point of view, we take the attitude of considering our membrane theories as IR limit of D2 brane configurations in type IIA in the presence of fluxes. This point of view has been emphasized in \cite{Aganagic:2009zk,Ami}. We motivate our construction by observing that all the involutions considered in this paper descend to an orientifold projection in type IIA.
From the field theory point of view,  we obtain the  ${\cal N}=1$ Chern-Simon theory  by projecting the parent Lagrangian.  As for ordinary orbifolds and orientifolds of four-dimensional gauge theories, we expect that planar equivalence holds  \cite{Lawrence:1998ja, Kakushadze:1998tr, Armoni:2008kr} and that the final theory will be  conformal when the parent theory  was.

Before finishing this introduction,  let us explain in more details the geometric quotient which is at the base
of all constructions in this paper.
Recall that 
%Calabi Yau manifold are well known to the physics comunity. 
a Calabi-Yau four-fold is a four dimensional complex manifold with a real (1,1) form $J$ and a complex (4,0) form 
$\omega$ satisfying
\begin{equation}
d J = 0\, , \qquad \qquad d \omega = 0
\end{equation}
$J$ is the Kahler form and $\omega$ the holomorphic four form. The manifold has holonomy $SU(4)$,
which is the group of transformations of the tangent bundle that leave $J$ and $\omega$ invariant.
On the other hand, Spin(7) manifolds are eight dimensional real manifolds with a globally defined self dual closed four form $\Omega_4$,
\begin{equation}
\Omega_4= * \Omega_4\, , \qquad\qquad d \Omega_4 = 0 \, .
\end{equation}
The manifold has holonomy Spin(7), which is the group of transformations of the tangent bundle that leave $\Omega_4$ invariant.

Every Calabi Yau four-fold $Y$ is also a Spin(7) manifold. Indeed, using  the Kahler form $J$ and the holomorphic volume form $\omega$, we can  define 
\begin{equation}
\Omega_4=\frac{1}{2} J \wedge J + \hbox{Re}(\omega)
\end{equation}
which is closed and self-dual. Of course the resulting Spin(7) manifold is not generic and actually it has holonomy $SU(4)$. To obtain a pure Spin(7) manifold we can consider an antiholomorphic involution $\Theta$ on Y.  It acts on $\omega$ and $J$ as: $\omega \rightarrow \bar{\omega}$, $J \rightarrow -J$ and therefore it does not commute with the complex structure. The four form $\Omega_4$ is invariant under $\Theta$ and the holonomy group is broken from $SU(4)$ to Spin(7). The quotient  $X=Y/\Theta$ is a Spin(7) manifold. 

In the rest of this paper we construct ${\cal N}=1$ Chern-Simons theories that naturally realize this construction
on their moduli space. We start in Section 2 by recalling the general structure of ${\cal N}=2$ Chern-Simons theories 
dual to Calabi-Yau four-folds. In Section 3 we discuss the orientifolding construction and, in Section 4, we provide
many examples of ${\cal N}=1$ Chern-Simons theories with Spin(7) moduli spaces $Y/\Theta$,  with quotient groups $\Theta$ with and without fixed points (except the tip of the cone). We privilege ${\cal N}=2$ Chern-Simons theories where the moduli space
is a complete intersection, or a simple set of algebraic equations in some ambient space,  but our construction is quite general. In Section 5 we discuss generalizations of the orientifold procedure to the case where gauge groups are identified; a large number of other  examples can be provided in this way.
We end with conclusions and comments.

\section{The ${\cal N}=2$ theories} 

There are by now several examples of world-volume theories for M2 branes probing four-dimensional Calabi-Yau singularities. The recent attitude is to consider ${\cal N}=2$ Chern-Simons  gauge theories \cite{Jafferis:2008qz,Martelli:2008si,Hanany:2008cd}. We consider quiver
theories with  $U(N)$ gauge groups and adjoint and bifundamental chiral matter superfields $X_{ab}$ interacting through a superpotential $W(X_{ab})$. There is no Yang-Mills kinetic term for the gauge groups but a Chern-Simons interaction with integer coefficients $k_a$, satisfying $\sum k_a =0$. The Lagrangian in ${\cal N}=2$ notations is reported in Appendix A.

In a standard ${\cal N}=2$ quiver with Yang-Mills interactions,  the moduli space is obtained by solving the F and D term constraints\footnote{We use the same symbol to denote the scalar ${\cal N}=2$ superfields and their lowest components.} 
\bea
\partial_{X_{ab}} W &=& 0\nonumber\\
\D_a(X)&\equiv & \sum_b X_{ab} X_{ab}^\dagger - \sum_c X_{ca}^\dagger X_{ca} + [X_{aa},X_{aa}^\dagger] \, =\, 0
\eea
and dividing by the gauge group. These kind of quivers are utilized in four dimensions to describe the superconformal world-volume theory of D3 branes probing Calabi-Yau three-fold conical singularities in type IIB.
In these cases the abelian moduli space, $Z$, is a Calabi-Yau three-fold. 

When the same quiver is used as a three dimensional theory for membranes,  the moduli space is bigger \cite{Jafferis:2008qz,Martelli:2008si,Hanany:2008cd}.
In ${\cal N}=2$ supersymmetry in three dimensions the gauge vector has a scalar partner $\sigma$, which, in a Chern-Simons theory,  has no kinetic term and  is  an auxiliary field.  The bosonic potential is
\beq \label{scabos2}
\sum_{X_{ab}} {\rm Tr} \left ( (\sigma _a X_{ab} - X_{ab} \sigma_b)(\sigma _a X_{ab} - X_{ab} \sigma_b)^\dagger  +|\partial_{X_{ab}} W|^2\right ) \, .
\eeq
where the auxiliary fields are determined by the constraints
\beq \D_a(X)= \frac{k_a}{2\pi} \sigma_a. \label{constraint}\eeq
The fields $\sigma_a$ can be eliminated  using these equations but sometimes we will  find convenient to keep them in the Lagrangian. The potential   is minimized  by
 \bea
\partial_{X_{ab}} W &=& 0\nonumber\\
 \sigma _a X_{ab} - X_{ab} \sigma_b &=& 0
 \label{eeqs}
\eea
In the abelian case all $\sigma_a\equiv\sigma$ are equal and the equations $ \D_a(X) =  \frac{k_a}{2\pi} \sigma$
reduce to the standard D terms of an ${\cal N}=2$ theory with a FI term depending on the Chern-Simons couplings.
Since $\sum_a k_a=0$ and $\sum_a \D_a(X)=0$ by construction, one of these equations is redundant. Moreover, any  linear combination of gauge groups with coefficient $m_a$ orthogonal to the CS parameters $\sum_a k_a
m_a=0$ has a vanishing moment map. We are thus imposing $g-2$ D-term constraints, where $g$ is
the number of gauge groups. We can impose simultaneously the D-term
constraints and the corresponding $U(1)$ gauge transformations by modding out by the complexified
gauge group. We do not need to impose the last D term condition  since it determines the value of the auxiliary field $\sigma$.  Moreover, the corresponding $U(1)$ group, through its CS coupling with the
overall gauge field, it is  broken to $\mathbb{Z}_k$, where $k=\gcd (\{k_a\})$ \cite{Distler:2008mk,Lambert:2008et,Aharony:2008ug}. As a result the abelian moduli space has a dimension of  one unit bigger than $Z$ and it has the general form  $Y/\mathbb{Z}_k$ \cite{Jafferis:2008qz,Martelli:2008si,Hanany:2008cd}. 

% It can be obtained from the solution of the abelian F term equations (the master space \cite{Forcella:2008bb, Forcella:2008eh}) by dividing by $G-2$ complexified $U(1)$ groups 
In general, $Y$ is  a $\mathbb{C}^*$ fibration over $Z$. $Z$ is uniquely specified by the gauge group and matter field of the quiver, while $Y$ is specified also by the choice of Chern-Simons couplings $k_a$: different values of the Chern-Simons levels correspond to different geometries Y. In the interesting case where $Z$ was a three-fold, we obtain a four-dimensional manifold. For toric quiver  based on Tilings, and some other generalization,  one can explicitly show that the moduli space is still Calabi-Yau \cite{Hanany:2008cd}.

\section{The orientifold construction and the ${\cal N}=1$ theories} 

%In this section we recall the construction of ${\cal N}=2$ Chern-Simons theories living on membranes at Calabi-Yau singularities, we show how to  obtain ${\cal N}=1$ theories corresponding to Spin(7) cones by an orientifold projection and we motivate the construction using branes.

%\subsection{}

Our plan is now to find  examples of theories probing Spin(7) cones $X=Y/\Theta$ by performing  a quotient on the ${\cal N}=2$ theory corresponding to the Calabi-Yau four-fold $Y$.  The anti-holomorphic involution $\Theta$ is a real orbifold of Y and we are lead to consider both orbifold and orientifold projections. 
In the case of M2 brane theories the orbifold projection is complicated by the fact that 
there is no open string description of M theory branes. In particular, even in the case of ${\cal N}=2$ holomorphic
abelian orbifolds, the standard rules for performing a quotient on  the Lagrangian give quivers whose moduli space
is slightly more complicated than the geometric abelian quotient \cite{Benna:2008zy,Berenstein:2009ay}. We now show that the field-theoretic orientifold procedure for
reducing the Lagrangian nicely reproduces Spin(7) cones of the required form. We first consider the cases where 
the number of gauge factors of the orientifolded theory is the same as in the parent theory. The case where
there is a further identification of gauge groups and matter fields will be discussed in Section 5.

We act on the Chern-Simons Lagrangian with an antiinvolution that conjugates the fields, thus breaking the
holomorphic structure and ${\cal N}=2$ supersymmetry. Here we summarize the transformations and
the results referring to the explicit examples in the next section for details. The projection will be
\bea
A^a_\mu && \rightarrow \,\,\,\,\,\,- \Omega_a  (A_\mu^a)^T \Omega_a^{-1} \nonumber\\
X_{ab}^i && \rightarrow \,\,\,\,\,\, \hbox{   } \hbox{   }\Omega_a  (X_{ab}^i)^* \Omega_b^{-1} \nonumber\\
\sigma_a  && \rightarrow \,\, \,\,\,\, \hbox{   } \hbox{   }\Omega_a  (\sigma_a)^T \Omega_a^{-1} 
\label{projection}\eea
or, when there is more than one bi-fundamental field $X_{ab}^i$ connecting the gauge groups $a$ and $b$, 
\beq
X_{ab}^i \rightarrow \,\,\,\,\,\, \eta_{ij} \Omega_a  (X_{ab}^j)^* \Omega_b^{-1} 
\eeq
where $\eta$ is a matrix satisfying $\eta^T \eta ={\rm Id}$.  The projection commutes only with the gauge transformations that satisfy $\Omega g^* = g \Omega$ and thus breaks the $U(2 N)$ groups to $O(2 N)$ for $\Omega=I$ and $USp(2 N)$ for $\Omega= J$, where $J$ is the $2N\times 2N$ symplectic matrix. The Chern-Simons
coupling  is $2k$ for $O(2N)$ and $k$ for $USp(2N)$ \cite{Aharony:2008gk, Lambert:2008et}.
We obtain consistency conditions for the projection of matter fields by taking the square of the transformation. In particular, when we have two different $X_{ab}^i$  connecting the group $a$ and $b$,  the choice $\eta_{ij}=\delta_{ij}$ (or a symmetric matrix of square one) requires $\Omega_a=\Omega_b$ and therefore same type of group, $O$ or $USp$, on the two sides, while $\eta_{ij}=\epsilon_{ij}$ requires   $\Omega_a\ne \Omega_b$ and two different groups in $a$ and $b$. 
%More details will be given when we analyze explicit examples and are summarized in Appendix B. 
The projection is required to leave the bosonic potential and the full Lagrangian invariant. Note in particular that
the D terms transform as $\D_a\rightarrow \Omega_a \D_a^T \Omega_a^{-1}$ consistently with the action on auxiliary fields and the constraint (\ref{constraint}). 

Since $\sigma_a$ are ${\cal N}=2$ superpartners of the gauge vector, it follows from the different transformations
of $A_\mu^a$ and $\sigma_a$ that the ${\cal N}=2$ supersymmetry is broken. This is also obvious from the fact that
the projection does not preserve the holomorphic structure of the Lagrangian. As explicitly shown in the
Appendix A, the projection preserves ${\cal N}=1$ supersymmetry. This follows from the decomposition of ${\cal N}=2$ into ${\cal N}=1$ supermultiplets. The vector multiplet decomposes in an ${\cal N}=1$ vector multiplet $\Gamma_a^{\alpha}$ and a real scalar superfield $R_a$ whose lowest component is $\sigma_a$. The chiral multiplets become ${\cal N}=1$  multiplets  $Y_{ab}$ %(with lowest component $X_{ab}$) 
each containing two real scalar superfields. The transformations (\ref{projection}) can be straightforwardly extended to an action on ${\cal N}=1$
superfields 
\bea
\Gamma_a^{\alpha} && \rightarrow\,\,  -\Omega_a (\Gamma_a^{\alpha})^T \Omega_a^{-1}\nonumber\\
R_a  && \rightarrow \,\, \,\,\,\,  \Omega_a  (R_a)^T \Omega_a^{-1} \nonumber\\
Y_{ab}^i && \rightarrow \,\,\,\,\,\, \eta_{ij} \Omega_a  (Y_{ab}^j)^* \Omega_b^{-1} 
\eea
that leaves the action invariant.

We are now ready to study the moduli space of the ${\cal N}=1$ theory in the case of one membrane, $N=1$.
The theory is a real quiver with $O(2)$ and $USp(2)\equiv SU(2)$ gauge groups. The projection identifies some
components of the original fields that  must now satisfy\footnote{We will use the symbol X to denote the ${\cal N}=2$ scalar superfields and their lowest scalar components.  We will use the symbol Y to denote the ${\cal N}=1$ scalar superfields and we will use it when we want to emphasize the ${\cal N}=1$ supersymmetry. The lowest component of the superfields $X$ and $Y$ is the same and it will be denoted $X$.}  
\beq X_{ab}^i = \eta_{ij} \Omega_a  (X_{ab}^j)^* \Omega_b^{-1} \label{projected}\eeq
The vacuum conditions are obtained  by projecting equations (\ref{eeqs}). We now make an ansatz for the vacuum configuration.  The fields satisfying  $X_{ab}= \Omega_a X_{ab}^* \Omega^{-1}_b$, which  connect the same type of gauge group, can be chosen in the form
\beq  X_{ab} = {\rm Re} (x_{ab}) I +  {\rm Im} (x_{ab}) J \label{fieldI}\eeq
where $x_{ab}$ is a complex number \footnote{The case   with symmetric $\eta_{ij}$ can be treated similarly}. 
Here and in the following $I$ and  $J= i\sigma_2$ denote the two by two identity and symplectic matrix, respectively. 
The fields  satisfying $X_{ab}^i =\epsilon_{ij}  \Omega_a (X_{ab}^j)^* \Omega_b^{-1}$, which connect an $O(2)$ to an $USp(2)$  gauge group, can be chosen in the form
\bea
X_{ab}^1 &=&  \frac{1}{\sqrt{2}} \left( ({\rm Re} (x_{ab}^1) I +  {\rm Im} (x_{ab}^1) J) +  i ( {\rm Re} (x_{ab}^2)  I +   {\rm Im} (x_{ab}^2) J )\right ) \nonumber\\
X_{ab}^2 &=&- \Omega_a(X_{ab}^1)^* \Omega_b^{-1}  \label{fieldII}
\eea
with similar expressions for fields connecting $USp(2)$ to $O(2)$ gauge groups. 
%where the plus and minus refers to fields connecting one $O(2)$ to one $SU(2)$ group and to fields connecting one $SU(2)$ to one $O(2)$ group, respectively.
%The fields  (\ref{fieldI}) typically connect gauge groups of same type, while  fields (\ref{fieldII}) connect one $O$  to one $USp$ group. 
In many cases one can show that these configurations exhaust the vacuum space. 
Note that, with the above ansatz,  we have the same number of complex fields and the same residual 
gauge symmetry as the original  ${\cal N}=2$ abelian quiver. Each gauge group is indeed broken to $U(1)$; the $SO(2)\subset O(2)$ and the real section $SO(2)\subset SU(2)$ leave the ansatz invariant. In all our examples, D terms
and F terms, restricted to the moduli space,   reduce to
\beq \D_a = \left (\sum_b |x_{ab}|^2 - \sum_c |x_{ca}|^2\right ){\rm Id} \, , \qquad\qquad \partial_{x_{ab}} W  =0 \, .\eeq
The auxiliary fields are then determined by $\sigma_a = 2\pi \D_a/ k_a$; the $\sigma_a$  are diagonal and the remaining equations $\sigma_a X_{ab} = X_{ab}\sigma_b$ are trivially satisfied. Moreover the residual $U(1)$ gauge symmetry acts on the fields $x_{ab}$  exactly as in  the parent ${\cal N}=2$ theory. In particular, 
$\sum_a k_a U(1)_a$ is broken to $\mathbb{Z}_{2k}$. 
At this point the moduli space would be $Y/\mathbb{Z}_{2k}$, similarly as the ${\cal N}=2$ case. However, we now have an extra discrete transformation acting on
the moduli space, obtained using the parity inversion $\sigma_3\in O(2)$ and the element   $i \sigma_3\in  SU(2)$. The transformation acts as
\beq  x_{ab} \rightarrow x_{ab}^* \eeq
on the fields (\ref{fieldI}), and
\bea
x_{ab}^1 && \rightarrow \hbox{  } \hbox{  }(x_{ab}^2)^* \nonumber\\
x_{ab}^2 && \rightarrow  - (x_{ab}^1)^* \nonumber\\
\eea
on the fields (\ref{fieldII}). This transformation preserves the vacuum conditions and can be lifted to an anti-involution $\Theta$ on $Y$. Combining with the $\mathbb{Z}_{2k}$ transformation we obtain a bigger
group that we call $\Theta_k$. The moduli space is $Y/\Theta_k$ and it is a Spin(7) cone.  

In the following we will construct many explicit examples of $\Theta_k$ groups acting without fixed points, except the origin of the  cone.

\subsection{Descending to type IIA}

We expect large quantum corrections to an ${\cal N}=1$ theory and its moduli space. 
We would like to have an explicit membrane realization of the previous construction, so that, by taking a near horizon geometry, we could argue for the superconformal invariance of the Chern-Simons theory and the form of the corresponding moduli space.

In order to obtain an open string description where we can explicitly compute the effect of an orbifold or orientifold projection we can use a duality with a type II string. 
%The original
%ABJM model and few of its generalizations can be described by a web of D3 and NS branes in type IIB.
%For a general chiral quiver there is no such description. 
A general construction\footnote{The original ABJM model and few of its generalizations can be described by a web of D3 and NS branes in type IIB.  However, for a general chiral quiver there is no such description in type IIB.}  involves shrinking the M theory circle and descending to type IIA. Various membrane theories have been explained in this way \cite{Aganagic:2009zk} and, although the construction does not explain all the proposed M2 theories and there are still few points to clarify, this construction  is a good laboratory where to test our theories. 

 We already observed that  that we can write Y as a  $\mathbb{C}^*$ fibration. More precisely, we can see $Y$ as a double fibration of the three dimensional CY $Z$ over a real line $\mathbb{R}$, with coordinate $\sigma$, and a circle $S^1$, with coordinate $\psi$.  The circle is the one acted upon by the discrete symmetry $\mathbb{Z}_{2k}$
and the one used  for descending from eleven to ten dimensions. In type IIA the ${\cal N}=2$ Chern-Simons  theory is realized on D2 branes probing a seven dimensional transverse space, which is $Z$ fibered over a line.  The Chern-Simons couplings on the D2 world-volume theory are induced by the RR fluxes generated by the dimensional reduction from M theory. The Chern-Simons couplings are assumed to dominate the IR physics and drive the D2 theory to an IR fixed point.  In this picture,  $\psi$ parametrizes the M theory circle and $\sigma$ is the value (FI) of the D term which is not imposed in the Chern-Simons theory. 

In the type IIA  description we are working with open strings and we can try to implement the  projection to ${\cal N}=1$ on D2 branes.  For this we need to understand how the action of $\Theta$ descends to type IIA. 
Locally the Calabi-Yau holomorphic four form $\omega$ will be given by
 $$\omega \sim f(z_i) \hbox{ }dz_1 \wedge dz_2 \wedge dz_3 \wedge (d\sigma + i d \psi)$$
where $z_i$ are local holomorphic coordinates for the three-fold $Z$.
 % and we obtain D2 branes probing  a transverse space which is the three dimensional Calabi-Yau  $Z$ fibered over a real line. We also obtain RR fluxes. In a simpler background $Z\times \mathbb{R}$ the theory on D2 branes
%would be the standard quiver Yang-Mills theory associated with $Z$. The presence of RR fluxes induces Chern-Simons
%couplings that dominates the IR physics and the theory flows to the expected ${\cal N}=2$ Chern-Simons ones. 
%In the type IIA  description we are working with open strings and we can try to implement the M theory orbifold.  
%We already observed that the four dimensional CY Y can be obtained by relaxing a D term in the D2 brane parent theory and ungauging the associated U(1). The FI of the D term $\sigma$ is the coordinate of the real line probed by the D2 brane, while the U(1) is the M theory circle with coordinate $\psi$.  
%We can indeed rewrite Y as a double fibration of the three dimensional CY $Z$ over a real line $\mathbb{R}$, with coordinate $\sigma$, and a circle $S^1$, with coordinate $\psi$. We can locally write the holomorphic four form as $$\omega_4 \sim f(z_i) z_1 \wedge z_2 \wedge z_3 \wedge (\sigma + i\psi)$$.
As we will discuss, all the  anti-holomorphic involutions $\Theta$ considered in this paper, which conjugate $\omega$ and change sign to $J$,  invert the sign of the M theory circle coordinate
\begin{equation}
 \psi \rightarrow -\psi \, 
\end{equation} 
and therefore  can be interpreted as orientifolds in type  IIA \cite{Sen:1996zq,Gopakumar:1996mu,Partouche:2000uq,Majumder:2001dx}.

The orientifold projection on D2 branes can be done with standard methods. Following this chain of dualities, we see that the ${\cal N}=1$ orientifold theory flows in the IR to the world-volume theory of membranes probing $Y/\Theta$.

%For some of the following examples, the previous argument can be made rigorous and prove that the proposed
%${\cal N}=1$ theories really correspond to Spin(7) cones.  

%Y is defined by the zero locuus of a set of polinomial $p_l$ in some complex coordinate $x_i$. We introduce a set of charges $Q_i$ associated to the relaxed U(1) gauge symmetry. The three dimensional cone is CY and the charges sum up to zero: $\sum Q_i =0$. Let us impose the symplectic quotient constraints:
%\begin{equation}
%\sum |x_i|^2Q_i = \sigma \qquad, \qquad x_i e^{iQ_i\psi} = x_i
%\end{equation}      
%the resulting manifold is three dimensional and it is a CY. Y is locally $\tilde{Y} \times S^1_{\psi} \times \mathbb{R}_{\sigma}$. We can consider the local coordinate $z_k$ of $\tilde{Y}$, $\sigma$ and $\psi$ as local coordinates of the four dimensional CY Y. 
%The D2 branes are probing this geometry. 

%\subsection{Planar equivalence}
%ZZZZZZZZZZZZ
%Abbiamo qualcosa da dire? certo che una sezione qui con questo titolo ci starebbe proprio bene.
%eh si...pero mi sa che non abbiamo molto da dire...pace!
%ZZZZZZZZZZZ
\section{Examples} 

In this section we provide many examples of orientifolded quivers with ${\cal N}=1$ supersymmetry
and Spin(7) moduli space, obtained as  a quotient of a Calabi-Yau by a discrete group. By abuse of language, we will
refer to actions that fix only the tip of the cone as  actions without fixed points. We will provide examples both 
 with and without fixed points. As discussed in the previous Section, the construction is quite general and can be applied to a large class of quivers. Here we privilege,
for simplicity, the cases where the Calabi-Yau can be written as a simple set of algebraic equations in some ambient space. Other cases can be handled with the well developed machinery of Tilings \cite{Hanany:2005ve,Franco:2005rj} and the master space \cite{Forcella:2008bb, Forcella:2008eh,Forcella:2008ng}.  We also privilege the case where the quotient acts without fixed points.

\subsection{Quivers with two groups}

\begin{figure}[h]
\centering
\includegraphics[scale=0.8]{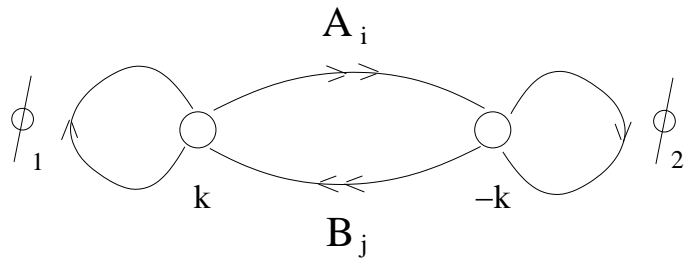}
\caption{The quiver for $V_{5,2}$, $\mathbb{C}^4$ and $C(T^{1,1})\times \mathbb{C}$.}
\label{twoF}
\end{figure}

We consider the simple ${\cal N}=2$ quiver theory studied in \cite{Martelli:2009ga} and presented in  Figure \ref{twoF}. It has two gauge groups $U(N)$, bi-fundamental fields $A_i$ and $B_j$ in the $(N,\bar{N})$ and $(\bar{N}, N)$, and  two adjoints $\phi_1,\phi_2$ interacting with the superpotential 
\begin{equation}
W= \frac{\phi_1^n}{n} + (-1)^{n+1} \frac{\phi_2^n}{n} + \phi_1 ( A_1 B_1 + A_2 B_2) + \phi_2 ( B_1 A_1 + B_2 A_2)
\end{equation}
 and Chern-Simons couplings $(k,-k)$.
This example is particularly simple because we can consider the fundamental fields in the Lagrangian as 
embedding coordinates of $Y$, which is described as a complete intersection in $\mathbb{C}^5$. 

\subsubsection{The cone over $V_{5,2}$}

Consider the cubic case, $n=3$, studied in detail in \cite{Martelli:2009ga}. To understand what is the corresponding
Calabi-Yau four-fold  we can study  the abelian case. The moduli space of the theory is   given by the solution of the F terms
\begin{equation}\label{modv52}
\phi_1 = - \phi_2 = \phi \qquad, \qquad \phi^2 + A_1 B_1 + A_2 B_2 = 0
\end{equation}
With a change of variable the second equation can be rewritten as
\begin{equation}\label{v52}
z_1^2 + z_2^2 + z_3^2 + z_4^2 + z_5^2 = 0
\end{equation}
which is the simplest example of Calabi Yau four-fold singularity, the quadric in $\mathbb{C}^5$.
It is the real cone over the $V_{5,2}$ Sasaki-Einstein 7 manifold. It is the obvious generalization of the  conifold singularity \cite{Klebanov:1998hh}. 

We still have to impose the D term constraints and mod by the gauge groups. 
One of the $U(1)$s acts trivially on matter fields and the corresponding D term vanishes.  The auxiliary fields $\sigma_1=\sigma_2=\sigma$ are equal and we remain with the  D term equation and gauge transformation
\begin{equation}
\sum_i \big|A_i\big|^2-\big|B_i\big|^2= \sigma\, ,  \qquad\qquad (A_i,B_i,\phi) \rightarrow (e^{i\psi}A_i,e^{-i\psi}B_i,\phi) 
\label{Dterm}
\end{equation}
However, as discussed in the previous Section, these conditions do not change the dimension of  the moduli space. 
In fact the first equation fixes the value of the auxiliary field $\sigma$ and the $U(1)$ action is broken to $\mathbb{Z}_k$ by the Chern-Simons interaction.  The moduli space is $C(V_{5,2})/\mathbb{Z}_k$.
Note that the equations (\ref{Dterm}) allow to think the four-fold CY as a double fibration of the conifold  over $S^1$ and $\mathbb{R}$ parametrized by $\psi$ and $\sigma$.  

We now construct orientifold models starting from the ${\cal N}=2$ theory. According to the choice of projection
we can have $O(2N)\times USp(2N)$ gauge groups (model I),  $O(2N)\times O(2N)$ gauge groups (model II)
and $USp(2N)\times USp(2N)$ gauge groups (model III).   We will explain in details the construction for model I. 
\\

\noindent {\bf Orientifold model I} 
 A natural antiholomorphic involution on the $\mathbb{C}^5$ complex coordinates that fixes only the origin of $Y=C(V_{5,2})$ is:
\begin{equation}\label{antiv52}
\phi \rightarrow   \phi^*\, ,  \qquad \qquad A_i \rightarrow \epsilon_{ij} A_i^*\, ,\qquad\qquad B_i \rightarrow \epsilon_{ij} B_i^* 
\end{equation}
It maps equations (\ref{modv52}) into their complex conjugate and therefore preserves $Y$. The action squares to
minus one and it is really a $\mathbb{Z}_4$ action. It is a $\mathbb{Z}_2$ action on $C(V_{5,2})/\mathbb{Z}_{2k}$. If we think of $Y$ as a fibration over a CY$_3$ as in equation (\ref{Dterm}), we see that 
the antiholomorphic action inverts the M theory angle $\psi$ and it corresponds to  an orientifold action in the IIA stringy realization of the M2 brane theory. Since the action has no fixed points except the origin, we can see it as a generalization of an O2 plane.   

We implement now the anti-involution (\ref{antiv52}) as an orientifold action on the ${\cal N}=2$ $U(2N)\times U(2N)$ quiver with Chern-Simons couplings $(2k,-2k)$.
We act as in (\ref{projection}) with $\Omega_1=I$ and $\Omega_2=J$ and $\eta_{ij}=\epsilon_{ij}$. Explicitly,   
the orientifold action on the quiver theory matter fields is
\begin{equation}
A_i \rightarrow - \epsilon_{ij} A_j^* J \, ,\qquad\qquad B_i \rightarrow  \epsilon_{ij} J B_j^* \, ,\qquad\qquad \phi_1 \rightarrow \phi_1^* \, ,\qquad\qquad \phi_2 \rightarrow  - J \phi_2^* J
\label{p1}
\end{equation}
and on the auxiliary fields is
\beq
\sigma_1\rightarrow (\sigma_1)^T\, , \qquad\qquad \sigma_2\rightarrow - J (\sigma_2)^T J \, . 
\eeq
The action is a symmetry of the Lagrangian. The projection send $W$ into its complex conjugate $W^*$ and the  bosonic potential is invariant. Moreover the D terms transforms as 
\beq D_1\rightarrow (D_1)^T \, , \qquad\qquad D_2\rightarrow -J (D_2)^T J\eeq
consistently with the auxiliary field constraints $D_a=\frac{k_a}{2\pi} \sigma_a$.  It is then easy to see that the full Lagrangian is invariant.  Note that  the above action square to one on the matter fields. Any other choice of $\Omega_{1,2}$ when 
$\eta_{ij}=\epsilon_{ij}$
would lead to an action that square to minus one and would project out the $A$ and $B$ fields. 

After orientifold projection, we obtain an ${\cal N}=1$ theory with gauge group $O(2N)\times USp(2N)$. 
The Lagrangian is obtained by restricting to the fields configurations invariant under (\ref{p1}). 
The Chern-Simons coefficients 
become $2k$ for $O(2N)$ and $k$ for $USp(2N)$. A reason for starting with coefficient $2k$ in the parent ${\cal N}=2$ theory is that the final coefficient for $USp(2N)$, which results to be $k$, must be an integer  \cite{Aharony:2008gk, Lambert:2008et}.

%\subsection{The moduli space}

%We would like check if the $\mathcal{N}=1$ CS field theory we proposed in the previous section is really the theory living on N M2 branes at the tip of the Spin(7) cone $C(V_{5,2})/\Theta$. 
As a consistency check we compute the moduli space. Let us consider the case of just one M2 brane. The gauge group reduces to $O(2) \times SU(2)$. The fields are two by two matrices. Let us consider the ansatz: 
\begin{eqnarray}\label{anstzv52}
& &A_1= \frac{1}{\sqrt{2}}\Big(\big( \hbox{Re}(a_1) +  \hbox{Im}(a_1) J \big) + i \big( \hbox{Re}(a_2)+ \hbox{Im}(a_2) J\big) \Big)\, \qquad  A_2=  A_1^* J \nonumber\\
& &B_1= \frac{1}{\sqrt{2}}\Big(\big( \hbox{Re}(b_1) +  \hbox{Im}(b_1) J \big) - i \big( \hbox{Re}(b_2)+ \hbox{Im}(b_2) J\big) \Big) \, \qquad  B_2=   -J B_1^*  \nonumber\\
& &- \phi_1 = \phi_2 = \hbox{Re}(\phi)+ \hbox{Im}(\phi) J
\end{eqnarray} 
with $a_1,a_2,b_1,b_2,\phi$ complex numbers. 
To find the vacua we can consider that the bosonic potential is the restriction of the ${\cal N}=2$ one.
In particular it is still a sum of squares minimized by 
\bea
\partial_{X_{ab}} W &=& 0\nonumber\\
 \sigma _a X_{ab} - X_{ab} \sigma_b &=& 0
\label{ch} 
\eea
where $\D_a(X)= \frac{k_a}{2\pi} \sigma_a$.  On the ansatz the D terms are
\beq \D_1=-\D_2 =  \left ( \sum_{i=1}^2 |a_i|^2 -|b_i|^2 \right )  {\rm Id} \eeq
It follows from $\D_a(X)= \frac{k_a}{2\pi} \sigma_a$ that $\sigma_1=\sigma_2$ are diagonal; the  second equation in (\ref{ch})
is then automatically satisfied. The F terms impose the complex constraint
\begin{equation}
a_1 b_1 + a_2b_2 + \phi^2 =0
\end{equation}  
These  is exactly the equation for the real cone over $V_{5,2}$.

 We would have obtained the same result by studying the Lagrangian in an ${\cal N}=1$ formalism
which makes manifest the residual supersymmetry. As discussed in Appendix A, the matter fields are combinations of real scalar multiplets interacting through an ${\cal N}=1$ real superpotential. After elimination of the auxiliary multiplets $R_a$, the superpotential expressed in terms of $A_1, B_1$ and $\phi_i$, reads
\bea
W_{{\cal N}=1}=&& \frac{2}{3} \phi_1^3 + \frac{2}{3} \phi_2^3 + 2 \phi_1 ( A_1 B_1 + A_1^* B_1^*) + 2 \phi_2 ( B_1 A_1 -J B_1^* A_1^* J) \nonumber\\
&&+ \frac{\pi}{k}\Big( \Big( A_1 A_1^{\dagger} + A_1^*A_1^T - B_1^{\dagger} B_1 - B_1^T B_1^*   +
\phi_1\phi_1^T  - \phi_1^T \phi_1 \Big)^2 \nonumber\\ 
&& -\Big( B_1 B_1^{\dagger}  - J B_1^*B_1^T J - A_1^{\dagger} A_1 + J A_1^T A_1^* J- \phi_2 J \phi_2^T J + J \phi_2^T J\phi_2 \Big)^2 \Big)
%\Big( A_i^{\dagger} A_i - B_i B_i^{\dagger} + \phi_2^{\dagger}\phi_2 - \phi_2 \phi_2^{\dagger}\Big)^2 - \Big( A_i A_i^{\dagger} - B_i^{\dagger} B_i + \phi_1 \phi_1^{\dagger} - \phi_1^{\dagger} \phi_1 \Big)^2\Big)  
\eea
The vacua are obtained by minimizing the real superpotential and it is easy to see that we re-obtain the above result.

There is a residual gauge symmetry on the moduli space. The $SO(2)$ groups contained in the $O(2)$ and the real subgroup $SO(2)\subset SU(2)$ factors preserve the ansatz. They act on the moduli space coordinates as
\bea
&&SO(2)\subset O(2)\, :  \qquad \qquad  a_i \rightarrow e^{i\psi_1} a_i\, ,\,\, b_i  \rightarrow e^{-i\psi_1} b_i\nonumber\\
&&SO(2)\subset SU(2)\, :  \qquad \qquad  a_i \rightarrow e^{-i\psi_2} a_i\, ,\,\, b_i  \rightarrow e^{i\psi_2} b_i
\eea
as in the parent ${\cal N}=2$ theory. One of the two groups acts trivially on fields, while the other is broken to $\mathbb{Z}_{2k}$ by the Chern-Simons interaction and acts as $(a_i,b_i^*)\rightarrow e^{i\pi/k}(a_i,b_i^*)$ while leaving $\phi$ invariant.  There is also a residual discrete symmetry  given by  the $\mathbb{Z}_2$ gauge symmetry acting as the inversion $\sigma_3$ inside the $O(2)$ and as $i\sigma_3$ inside $SU(2)$.  Its action on (\ref{anstzv52}) reproduces exactly the antiholomorphic involution (\ref{antiv52}),
\begin{equation}
\phi \rightarrow \phi^* \qquad,\qquad a_i \rightarrow \epsilon_{ij}a_j^*\qquad,\qquad b_i \rightarrow \epsilon_{ij} b_j^*
\end{equation}
The final result for the moduli space is $C(V_{5,2})/\Theta_k$, where $\Theta_k$ is the discrete group generated
by $\mathbb{Z}_{2k}$ and the anti-involution.
\\

\noindent {\bf Orientifold  model II and III: } we could have chosen  anti-involutions with fixed points.
The simplest one would send
\beq a_i\rightarrow a_i^*\, ,\qquad\qquad b_i\rightarrow b_i^* \, , \quad\qquad \phi_i\rightarrow \phi_i^* \label{realv52}\eeq
It leaves the equation for $C(V_{5,2})$ invariant but has a fixed point locus of real dimension four.
The transformation can be implemented on the ${\cal N}=2$ theory as
\beq A_i\rightarrow \Omega A_i^* \Omega^{-1} \, ,\qquad\qquad B_i\rightarrow \Omega B_i^* \Omega^{-1} \, , \qquad\qquad \phi_i\rightarrow \Omega \phi_i^* \Omega^{-1}\eeq 
with $\Omega=I$ or $\Omega=J$. The resulting ${\cal N}=1$ theories have $O(2N)\times O(2N)$ or $USp(2N)\times USp(2N)$ gauge groups. It is easy to see that the moduli space in the case of one membrane, $N=1$, correspond
to the quotient of $C(V_{5,2})$ by a discrete group obtained by combining the usual abelian projection with
the real involution (\ref{realv52}). Note that the  two types of theories with orthogonal or symplectic groups
have the same moduli space. A similar phenomenon happens for D3 branes in type IIB \cite{Witten:1998xy}.
 The action descends to type IIA as an orientifold projection that fixes a six-dimensional plane and generalize an O6 plane. The two types of theories  presumably correspond to two types of O6 action with the same geometrical lift to M theory.

\subsubsection{The case of $\mathbb{C}^4$}
In the case $n=2$ we can integrate out the adjoint fields and obtain a quiver with fields $A_i$ and $B_j$
interacting with the ${\cal N}=2$ superpotential
\beq W =\epsilon_{ij}\epsilon_{pq} A_i B_p A_j B_q \eeq
This is the original ABJM model \cite{Aharony:2008ug} and describe membranes on $\mathbb{C}^4/\mathbb{Z}_k$. The complex coordinates $A_{i}$, $B_{j}$ parametrize $\mathbb{C}^4$ and the $\mathbb{Z}_k$  action is $(A_{i},B_{j}) \rightarrow ( e^{2\pi i/k}A_{i},e^{-2\pi i/k}B_{j})$. 

We can consider a simple anti-involution of $\mathbb{C}^4$ without fixed points 
\begin{equation}
A_1 \rightarrow  A_2^* \qquad A_2 \rightarrow  -A_1^* \qquad B_1 \rightarrow B_2^* \qquad B_2 \rightarrow  - B_1^*
\label{ABJMp1}
\end{equation}
and try to implement it as an orientifold projection on the ${\cal N}=2$ theory as we did for $V_{5,2}$
\begin{equation}
A_i \rightarrow - \epsilon_{ij} A_j^* J \, ,\qquad\qquad B_i \rightarrow  \epsilon_{ij} J B_j^* \label{ABJMp11}\end{equation}
The resulting theory has $O(2N)\times USp(2N)$ gauge groups. 
%The $\mathcal{N}=1$ ABJM superpotential is 
%{\bf serve??} 
%\begin{equation}
%W^{ABJM}_{\mathcal{N}=1}= \hbox{Tr} \big(\frac{1}{2L} \big( \big( A^{\dagger}_{i} A_{i} -  B_i B_i^{\dagger} \big)^2 - \big( A_{i} A^{\dagger}_{i} -  B^{\dagger}_i B_i \big)^2 \big) + \frac{1}{2k} \big( \epsilon_{ij} \epsilon_{pq} \big( A_{i} B_{p} A_{j} B_{q} +  A^*_{i} B^*_{p} A^*_{j} B^*_{q} \big)\big)\big) 
%\label{abjmW1} 
%\end{equation}
Repeating the same analysis as done above
for $V_{5,2}$ we will see that the one membrane moduli space is $\mathbb{C}^4/{\cal D}_k$ where ${\cal D}_k$
is a dihedral group  obtained by combining the $\mathbb{Z}_{2k}$ action with the antiinvolution \cite{Morrison:1998cs, Hanany:2008qc}. 

${\cal D}_k$ actually preserves a bigger ${\cal N}=5$ supersymmetry. The final Lagrangian has a hidden
${\cal N}=5$ supersymmetry and an $USp(4)$ global symmetry. The orientifold theory was indeed already constructed and discussed in \cite{Hosomichi:2008jb,Aharony:2008gk}. The ABJM model has 
a hidden ${\cal N}=6$ supersymmetry and $SU(4)$ global symmetry rotating $(A_1,A_2,B_1^\dagger,B_2^\dagger)$.
With an $SU(4)$ transformation we can map the  projection (\ref{ABJMp11}) into
\begin{equation} A_i \rightarrow -\epsilon_{ij} B_j^T J \, ,\qquad\qquad B_i \rightarrow  \epsilon_{ij} J A_j^T 
\label{ABJMp2}
\end{equation} 
where is manifest that at least an ${\cal N}=2$ supersymmetry is preserved.
 
We could obtain other real quotient of $\mathbb{C}^4$ with fixed points. For example
\begin{equation}
A_i \rightarrow  A_i^* \qquad B_i \rightarrow  B_i^* 
\end{equation}
is obtained by considering $O(2N)\times O(2N)$ or $USp(2N)\times USp(2N)$ orientifolds. 
%The supersymmetry
%is now ${\cal N}=1$.
 
\subsubsection{The case of the conifold times $\mathbb{C}$}

The simplest model has $n=0$ \cite{Hanany:2008cd}. In the ${\cal N}=2$ theory a single membrane probes a moduli space given by
\beq A_1 B_1 +A_2 B_2 =0 \eeq
with the $\phi$ unrestricted. The theory has various branches. As usual in the study of quivers, we restrict
to the case $\phi_1=\phi_2$. We thus obtain a Calabi-Yau four-fold which is a conifold times $\mathbb{C}$  \cite{Hanany:2008cd}.
This cone is less interesting since it has a line of singularities. However we can still perform on the complex coordinates the anti-involution
\begin{equation}
A_1 \rightarrow  A_2^* \qquad A_2 \rightarrow  -A_1^* \qquad B_1 \rightarrow B_2^* \qquad B_2 \rightarrow  - B_1^* \qquad \phi_i\rightarrow \phi_i^*
\label{conp1}
\end{equation}
which acts with fixed points. 

We implement on the quiver fields the orientifold projection as above
\begin{equation}
A_i \rightarrow - \epsilon_{ij} A_j^* J \, ,\qquad\qquad B_i \rightarrow  \epsilon_{ij} J B_j^* \, ,\qquad\qquad \phi_1 \rightarrow \phi_1^* \, ,\qquad\qquad \phi_2 \rightarrow  - J \phi_2^* J
\end{equation}
obtaining an $O(2N)\times USp(2N)$ theory. 
%Using again the  ansatz (\ref{ansatzv52}) for a single membrane, the vaccum conditions reduces to
%\beq a_1 b_1 + a_2 b_2 =0 \eeq
%with unrestricted $\phi_i$. The residual gauge symmetry $SO(2)\times SO(2)$ is broken by the Chern-Simons interactions to $\mathbb{Z}_{2k}$. The extra discrete symmetry acting as the inversion $\sigma_3$ inside the $O(2)$ and as $-i\sigma_3$ inside $SU(2)$ gives
%\begin{equation}
%a_1 \rightarrow  a_2^* \qquad a_2 \rightarrow  -a_1^* \qquad b_1 \rightarrow b_2^* \qquad b_2 \rightarrow  - b_1^*
%\end{equation}
Repeating the same argument as above, we see that  the single membrane moduli space is $C(T^{1,1})\times \mathbb{C}$ divided by the discrete group generated by the anti-involution and the $\mathbb{Z}_{2k}$ action. 

\subsection{Anti-involution on $Q^{1,1,1}$ and its orbifolds}

\begin{figure}[h]
\centering
\includegraphics[scale=0.8]{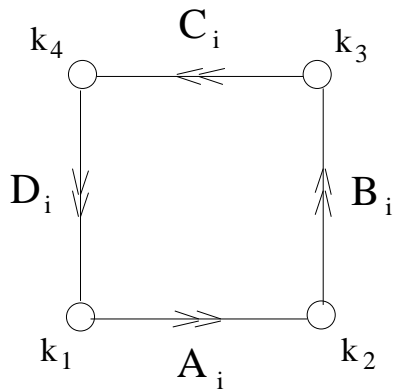}
\caption{This is the quiver for $Q^{1,1,1}$  when the Chern-Simons couplings are $(0,1,0,-1)$ and for $Q^{2,2,2}$ when $(1,1,-1,-1)$.}
\label{F0}
\end{figure}

The cone over the manifold $Q^{1,1,1}$ and its $\mathbb{Z}_2$ orbifold $Q^{2,2,2}$ can be obtained by considering
the chiral quiver in Figure \ref{F0}
with superpotential $W=\epsilon_{ij}\epsilon_{pq} A_i B_p C_j D_q$. 
The associated three-fold $Z$ is the complex cone over $F_0$ and it is a $\mathbb{Z}_2$ quotient of the
conifold. $Z$ is relevant in the application of the quiver to D3 branes theories in type IIB. Here we consider
the three-dimensional Chern-Simons theory with couplings $(k_1,k_2,k_3,k_4)$ with $k_1+k_2+k_3+k_4=0$.
The moduli space is a Calabi-Yau four-fold (a different one for each choice of  $k_i$)  which is a fibration over $Z$. In the special case $(0,1,0,-1)$ we obtain $Q^{1,1,1}$ \cite{Aganagic:2009zk} and in the case $(1,1,-1,-1)$ we obtain $Q^{2,2,2}$ \cite{Franco:2009sp,Davey:2009sr,Amariti:2009rb}.
This model can be analyzed by descending in type IIA. We would obtain D2 branes transverse to a seven-dimensional
manifold which is a fibration of $Z$ over a real line. 

\subsubsection{The cone over $Q^{1,1,1}$}

This case corresponds to the Chern-Simons couplings $(0,k,0,-k)$. The  abelian F terms are 
\begin{equation}
A_1 C_2 = A_2 C_1 \, , \qquad\qquad B_1 D_2 = B_2 D_1 \label{aft}
\end{equation}
The space of solution of the abelian F terms is called the master space \cite{Forcella:2008bb,Forcella:2008eh,Forcella:2008ng} and it is useful to understand the M2 branes moduli space \cite{Hanany:2008cd}. Here it has complex dimension six and it is  the product of two conifolds. 

To obtain the $CY_4$ from the six dimensional master space we need to impose the D term constraints and divide by the gauge groups. The relations $D_a =  \frac{k_a}{2\pi} \sigma_a$ teach us that we need to  impose the D term for the two gauge groups with zero Chern-Simons couplings and we need to mod by the corresponding $U(1)$. As usual,
we can dispose of both operations by dividing by the complexified gauge group. We thus quotient by
the two complexified gauge groups with zero Chern-Simons. One is acting on $C$ with charge $+1$ and on $B$ with charge $-1$ and the other is acting on $A$ with $+1$ and on $D$ with $-1$.  It is easy to get an algebraic description: the gauge invariants are
\begin{equation}
w_{ij} = B_i C_j \, , \qquad\qquad  W_{ij} = D_i A_j 
\end{equation}
and satisfy nine quadratic constraints
\begin{eqnarray}
&& w_{11} w_{22}= w_{12} w_{21}  \, , \qquad  W_{11} W_{22}= W_{12} W_{21} \nonumber \\
&& W_{i1} w_{j2}= W_{i2} w_{j1}  \, ,  \qquad  W_{1i} w_{2j}= W_{2i} w_{1j}   \,\,\, \qquad    i,j=1,2
\label{eqs}
\end{eqnarray}
where the first line comes from the definition of $w$ and $W$ and the second line from the F term equations. One of the eight equations in the second line is linearly dependent. Thus we obtain the description of $Y$ as a set of nine quadrics in $\mathbb{C}^8$. These equations define the cone over $Q^{1,1,1}$ \cite{Fabbri:1999hw}.

The remaining non-trivial  gauge group acts as
\begin{equation}
U(1)_2 - U(1)_4: \,\,\,\,  \qquad \left (\begin{array}{cccc}
A & B& C&D \\ 
\hline
-1&1&1&-1
\end{array}\right ) 
\label{U1}
\end{equation}
and it is broken to $\mathbb{Z}_k$ by the Chern-Simons interaction. The moduli space is then $C(Q^{1,1,1})/\mathbb{Z}_k$. Note that the $\mathbb{Z}_k$ action breaks the $SU(2)^3$ isometry of $Q^{1,1,1}$ to $SU(2)^2\times U(1)$ which
is precisely the symmetry of the quiver.

By modding by $U(1)_2 - U(1)_4$ 
we would obtain the complex cone over $F_0$. We can thus  see $Q^{1,1,1}$ as a $\mathbb{C}^*$ fibration over $F_0$. The leaves are copies of (resolutions of) $F_0$
\begin{equation}
\{ \sum_{i=1}^2 |B_i|^2+|C_i|^2 - |A_i|^2-|D_i|^2 = \sigma  \} /U(1)_2 - U(1)_4
\label {Dterm2}
\end{equation}
and we can choose local coordinates giving a point in $F_0$ and the value of $\sigma$ and
of the angle $\psi$ parametrizing the $U(1)$ orbit.

\vskip 0.5 truecm 

%We can define various anti-involutions of $Q^{111}$:

%\begin{itemize}

%{\item I. The obvious one is  $w_{ij}\rightarrow \bar W_{ij}$ which leaves invariant the nine quadratic equations
%defining $Q^{111}$. It can be realized as 
%\begin{equation}
%A_i \rightarrow \bar C_i \, \qquad \qquad B_i \rightarrow \bar D_i
%\end{equation}
%on the elementary fields. On $F_0$, it acts as an anti-involution. We see from eq. (\ref{Dterm}) that it acts
%by $\sigma\rightarrow -\sigma$ and, from eq. (\ref{U1}), by $\theta\rightarrow \theta$. By descending in type IIA we obtain a combined orbifold action on $F_0\times \mathbb{R}$ acting as an antiinvolution
%on $F_0$ and a reverse in sign of the coordinate $\sigma$ parameterizing $\mathbb{R}$.}
%{\item II. Another action just exchange the indices $i=1,2$ and perform a complex conjugation on $w$ and
%$W$ separately. For example, $w_{11}\rightarrow \bar w_{22}, w_{12}\rightarrow \bar w_{21}, \, ...$. 
%It also leaves invariant the nine defining quadratic equations.
%It
%correspond to
%\begin{equation}
%A_1\rightarrow \bar A_2 \, \qquad   B_1\rightarrow \bar B_2 \, \qquad   C_1\rightarrow \bar C_2 \, \qquad   D_1\rightarrow \bar D_2 \, \qquad   
%\end{equation}
%This action descends to an anti-involution of $F_0$. This time $\sigma\rightarrow\sigma$ while
%$\theta\rightarrow -\theta$. In type IIA we obtain an orientifold of  $F_0\times \mathbb{R}$.}
%
%\end{itemize}

%Both anti-involution act without fixed points.  

\noindent  {\bf The orientifold projection:} 
A possible anti-holomorphic action without fixed point on $Q^{1,1,1}/\mathbb{Z}_k$ is:
\begin{eqnarray}\label{inv}
& &  W_{11} \rightarrow \hbox{  }\hbox{   }W_{21}^* \qquad  W_{12} \rightarrow  \hbox{  }\hbox{   }\hbox{  } W_{22}^*\qquad w_{12} \rightarrow \hbox{  }\hbox{   } w_{22}^* \qquad w_{11} \rightarrow \hbox{  }\hbox{   } w_{21}^* \nonumber\\
& &  W_{21} \rightarrow - W_{11}^* \qquad  W_{22}\rightarrow - W_{12}^* \qquad w_{22} \rightarrow - w_{12}^* \qquad w_{21} \rightarrow - w_{11}^* \nonumber\\
\end{eqnarray}
It is actually a $\mathbb{Z}_4$ action, however it is a $\mathbb{Z}_2$ action on $Q^{1,1,1}/\mathbb{Z}_{2k}$. 
%It squares to one on $Q_{111}/\mathbb{Z}_k$ if k is even and different from zero.
The action on the elementary coordinates is, 
\begin{equation}\label{elfieldinv}
D_i \rightarrow  \epsilon_{ij} D_j^* \qquad B_i \rightarrow  \epsilon_{ij} B_j^* \qquad A_i \rightarrow A_i^* \qquad C_i \rightarrow  C_i^*
\end{equation}
This anti-holomorphic projection breaks the $SU(2)^2$ symmetry of the quiver theory to $SU(2) \times U(1)$. 
If we think of the four dimensional CY $C(Q^{1,1,1})/\mathbb{Z}_k$ as a double fibration of the complex cone over $F_0$ over $\mathbb{R}$ and $S^1$ as in (\ref{Dterm2}), 
%Let us introduce the real variables $\sigma$ and $\phi$, the leaves of the fibration are resolutions of $\mathbb{F}_0$ given by:
%\begin{equation}
%\{ \sum_{i,j}^2 |w_{ij}|^2 - |W_{ij}|^2 = \sigma  \} /U_{CS}(1)
%\label {DtermwW}
%\end{equation}
%
%where (\ref{elfieldinv})$U_{CS}(1)$ acts as $(w_{ij},W_{ij}^*) \rightarrow e^{i\phi}(w_{ij},W_{ij}^*)$. We choose as local coordinate the complex coordinates of $\mathbb{F}_0$ plus $\sigma$ and $\phi$. 
the projection (\ref{inv}) acts as an antiholomorphic involution on $F_0$, it leaves $\sigma$ invariant and it changes the sign of $\psi$: $\psi \rightarrow -\psi$. 
%We would like to obtain the $\mathcal{N}=1$ theory dual to this Spin(7) manifold. It is obvious that the involution \ref{elfieldinv} it is not consistent with the $U(N)^4$ gauge groups. It means that something must happen to the gauge groups. 
We can understand better this projection by descending in  type IIA. In these coordinates, the type IIA limit corresponds to shrink the circle associated to $\psi$ \cite{Aganagic:2009zk}. Following \cite{Sen:1996zq} the flip in the M theory circle will translate to an orientifold projection in type IIA.  

%Since the antiholomorphic involution (\ref{elfieldinv}) acts without fixed point on the four 
%dimensional CY,  
%and we do not expenct any identifications among the gauge fields. For this reason we aspect that the Orbifold projection will give $O(M)$ or $USp(M)$, but not $U(M)$ guauge groups. It acts as $\Gamma_{\alpha} \rightarrow -\Omega_i (\Gamma^{\alpha}_i)^T \Omega_i^{-1}$ on the i-th gauge supermultiplet, where $\Omega_i$ is the 2N dimensional Identity matrix for the ortogonal group and the 2N dimensional symplectic matrix J for the symplectic group. 

%The orientifold action on the matter fields is given by the antinvolution (\ref{elfieldinv}) dressed with an action $X_{ab} \rightarrow \Omega_a (X_{ab}') \Omega_b^{-1}$ on the gauge indices. 
We now realize the anti-holomorphic action as an orientifold projection on the Lagrangian.
The final gauge groups are of type $O(2N)$ or $USp(2N)$.  Consistency of the projection (\ref{elfieldinv}) requires that gauge groups 1 and 2 and gauge groups 3  and 4 are of same type, but the groups 1,2 are different from the groups 3,4. One possible choice is: O, O, USp, USp. The orientifold  action is then,
\begin{equation}\label{orientielfield}
D_i \rightarrow  \epsilon_{ij} J D_j^* \qquad B_i \rightarrow  -\epsilon_{ij} B_j^* J \qquad A_i \rightarrow  A_i^* \qquad C_i \rightarrow  -J  C_i^* J
\end{equation}
It is easy to check that this orientifold action is a symmetry of the original $\mathcal{N}=2$ theory. 
%We can then consider its quotient. 
%We believe that the $\mathcal{N}=1$ field theory we obtain with this procedure is the field theory dual of the Spin(7) manifold realized as the antiholomorphic quotient of the cone over $Q_{111}/\mathbb{Z}_k$ with respect to (\ref{inv}). 
The resulting field theory has O(2N)$\times$O(2N)$\times$USp(2N)$\times$USp(2N) gauge group and ${\cal N}=1$ supersymmetry. 
%To check that the proposed $\mathcal{N}=1$ CS matter theory  is really the dual theory of the Spin(7) manifold $C(Q_{111}/\mathbb{Z}_k)/\Theta$ 

To test the proposal, we compute the moduli space for one membrane. The gauge group is $O(2)\times O(2) \times SU(2) \times SU(2)$.  We parametrize elementary fields as,
\begin{eqnarray}
& & A_i = Re(a_i) + J Im(a_i)\, ,  \qquad  C_i =  Re(c_i) + J Im(c_i) \nonumber\\
& &B_1= \frac{1}{\sqrt{2}}\Big(\big( \hbox{Re}(b_1) +  \hbox{Im}(b_1) J \big) + i \big( \hbox{Re}(b_2)+ \hbox{Im}(b_2) J\big) \Big) \, \qquad  B_2=   B_1^* J \nonumber\\
& &D_2= \frac{1}{\sqrt{2}}\Big(\big( \hbox{Re}(d_2) +  \hbox{Im}(d_2) J \big) + i \big( \hbox{Re}(d_1)+ \hbox{Im}(d_1) J\big) \Big) \, \qquad  D_1=   J D_2^* 
\end{eqnarray}
where $a_i$, $b_p$, $c_j$, $d_q$ are generic complex numbers. The four $SO(2)$ gauge subgroups of the gauge groups act as the $U(1)^4$ gauge group of the unorientifolded theory.
The D term of the four gauge groups are proportional to the identity
\bea \D_1= \sum_i \left ( |a_i|^2 -|d_i|^2\right ) {\rm Id} \, , \qquad\qquad  \D_2= \sum_i \left ( |b_i|^2 -|a_i|^2\right ) {\rm Id}
\nonumber\\
 \D_3= \sum_i \left ( |c_i|^2 -|b_i|^2\right ) {\rm Id}\, , \qquad\qquad  \D_4= \sum_i \left ( |d_i|^2 -|a_i|^2\right ) {\rm Id}
 \eea   
The vacuum conditions require $\sigma_a X_{ab}=X_{ab} \sigma_b$  with $D_a =  \frac{k_a}{2\pi} \sigma_a$. The first condition
is trivially satisfied while the second one requires
\begin{equation}
\sum _i |a_i|^2=\sum_i |d_i|^2\, , \qquad\qquad \sum_i |c_i|^2=\sum_i |b_i|^2
\end{equation}
which are the D term of the $\mathcal{N}=2$ theory.  We should also mod by the $SO(2)$ subgroups of the first and third gauge groups which has zero Chern-Simons coupling, exactly as in the parent theory.
The F term equations give
\begin{equation}
a_1 c_2 = c_1 a_2\, , \qquad\qquad b_1 d_2 = b_2 d_1
\end{equation}
On top of these equations the CS interaction breaks the third $SO(2)$ gauge group to a  $\mathbb{Z}_{2k}$ action: $(a_i^*,b_p,c_j,d_p^*) \rightarrow e^{{ \pi i}/k} (a_i^*,b_p,c_j,d_p^*)$. Taking the quotient with respect to this discrete action we obtain exactly $C(Q^{1,1,1})/\mathbb{Z}_{2k}$. There is an extra discrete $\mathbb{Z}_2$   symmetry on the moduli space  acting as $\sigma_3$ in the $O(2)$ gauge groups and as $i \sigma_3$ in the $SU(2)$ gauge groups. Its action on the coordinates of the moduli space is
\begin{equation}
(a_i,c_i) \rightarrow (a_i^*,c_i^*)\, , \qquad\qquad (b_i,d_i) \rightarrow \epsilon_{ij} (b_j^*,d_j^*)
\end{equation}
which it is exactly the orientifold action we considered in the previous section. 
As a result,  the moduli space is $C(Q^{1,1,1})$ divided by the discrete group generated by the anti-involution and the $\mathbb{Z}_{2k}$ action.
\\

\noindent{\bf The ${\cal N}=1$ description:} We could have alternatively studied the moduli space using 
the ${\cal N}=1$  superfield formalism. 
The original ${\cal N}=2$ theory, written in ${\cal N}=1$  formalism,  contains four extra adjoint real superfields
$R^a$. The original fields $A_i,B_i,C_i,D_i$ are now considered as complexified ${\cal N}=1$ superfields.
As discussed in the Appendix, the  Chern-Simons contribution   is (see (\ref{CS2}))
\begin{equation}
\frac{k}{4\pi} \int d^2\theta_1 \big( R^2_4 - R^2_2 \big)  
\label{CS2}
\end{equation}
There is no quadratic term for $R_1$ and $R_3$ since the corresponding Chern-Simons couplings are zero. Considering the linear interactions (\ref{cin2}) and using the equation of motion for $R_1$, $R_2$, $R_3$, $R_4$, we obtain the constraints
\begin{equation}
A_iA_i^{\dagger} = D_i^{\dagger}D_i \qquad C_iC_i^{\dagger} = B_i^{\dagger}B_i
\label{con}
\end{equation}
and the following contribution to the $\mathcal{N}=1$ superpotential coming from the D terms,
\begin{equation}
\frac{\pi}{k}\int d^2\theta_1 \big(  B_i B_i^{\dagger} -A^{\dagger}_{i} A_{i} \big)^2 - \big( D_i D_i^{\dagger} - C^{\dagger}_{i} C_{i} \big)^2
\label{c}
\end{equation}
which should be added to the contribution coming from the ${\cal N}=2$ superpotential
\begin{equation}
\int d^2\theta_1 \epsilon_{ij} \epsilon_{pq} \hbox{ Tr } \big( A_{i} B_{p} C_{j} D_{q} +  A^*_{i} B^*_{p} C^*_{j} D^*_{q} \big)\, .
\label{p}
\end{equation}

The ${\cal N}=1$ theory can be obtained by restricting the ${\cal N}=1$ superpotential to configurations invariant under  (\ref{orientielfield}). We can express the fields in terms of the independent ones  $A_i,C_i,B_1,B_1^*,D_1,D_1^*$, where
$B_1$ and $D_1$ are arbitrary complex matrices while $A_i$ and $C_i$ satisfy the reality conditions $A_i=A_i^*$ and $C_i=-JC_i^* J$. 
The $\mathcal{N}=1$ superpotential comes from (\ref{con},\ref{c},\ref{p}) and reads
\begin{eqnarray}
W_{{\cal N}=1}&=& \frac{\pi}{k} \big(\big( B_1 B_1^{\dagger} + B_1^* B_1^{T} - A^{T}_{i} A_{i}\big)^2 - \big( D_1 D_1^{\dagger} - J D_1^* D_1^{T}J + J C^{T}_{i} J C_{i}\big)^2\big) \nonumber\\
& &- 2 \epsilon_{ij} \hbox{ Tr } \big( A_i B_1 C_j J D_1^* +  A_i B_1^* J C_j D_1 \big) \nonumber\\
& & \alpha (A_iA_i^{T} - D_1^{\dagger}D_1 - D_1^{T}D_1^* ) + \beta (- C_i J C_i^{T}J - B_1^{\dagger}B_1 +J B_1^{T} B_1^* J) \nonumber\\
\label{pinv}
\end{eqnarray}
with $\alpha, \beta$ two Lagrange multipliers.
The vacua can be obtained by minimizing the ${\cal N}=1$ superpotential. The result agrees with that described 
above.

\subsubsection{The cone over $Q^{2,2,2}$}

$Q^{1,1,1}$ admits a supersymmetric orbifold $Q^{2,2,2}$ that reduces by half the length of the $U(1)$ fiber. The field theory living on N M2 branes at the tip of the cone over $Q^{2,2,2}$ is described by the same quiver and superpotential of the $Q^{1,1,1}$ theory, but it has different values for the CS levels $(1,1,-1,-1)$ \cite{Davey:2009sr,Franco:2009sp}.  

Consider as usual the abelian case. One $U(1)$ gauge group acts trivially on matter fields and another one is broken by the Chern-Simons interactions. The surviving $U(1)$ actions
\begin{equation}
 \left (\begin{array}{cccc}
A & B& C&D \\
\hline 
1&0&-1& 0\\
2&-1&0&-1
\end{array}\right ) 
\label{U1q222}
\end{equation}
can be paired with
\beq \D_1+ \D_4 = \sum_{i=1}^2 |A_i|^2 - |C_i|^2 = 0 \qquad  \D_1 -\D_2 =  \sum_{i=1}^2 2 |A_i|^2 - |B_i|^2 - |D_i|^2 = 0 \eeq
(following from $\D_a=\frac{k_a}{2\pi} \sigma_a$) to give two complexified $U(1)$. There are  $27$ invariants 
\begin{equation}
\alpha_{ij\, pq} =A_i B_p C_j D_q\, , \qquad\qquad \beta_{ij\, pq}= A_i B_p C_j B_q\, , \qquad\qquad \gamma_{ij\, pq} =A_i D_p C_j D_q
\end{equation}
totally symmetrized in the indices $i,j$ and $p,q$ by the F term equations (\ref{aft}). They satisfy $253$ quadratic 
equations due to the F terms and give an algebraic description of $C(Q^{2,2,2})$ as a set of quadrics in $\mathbb{C}^{27}$. For generic $k$ the moduli space is $C(Q^{2,2,2})/\mathbb{Z}_k$ with $\mathbb{Z}_k$ acting as
\beq B_i\rightarrow e^{\frac{i\pi}{k}} B_i\, , \qquad\qquad D_i\rightarrow e^{-\frac{i\pi}{k}} D_i\, . \eeq

A natural antiholomorphic involution on $C(Q^{2,2,2})$ elementary coordinate fields is
\begin{equation}\label{elfieldinv22}
A_i \rightarrow  i \epsilon_{ij} A_j^* \qquad B_i \rightarrow  B_i^* \qquad C_i \rightarrow -i \epsilon_{ij} C_j^* \qquad D_i \rightarrow  D_i^* 
\end{equation}
It is an easy exercise to check that this action, which fixes a subspace of $\mathbb{C}^{27}$, has however no fixed points on the four-fold.

The ${\cal N}=1$ Chern-Simons theory  $C(Q^{2,2,2})/\Theta_k$ where $\Theta_k$ is
the semi-direct product of the anti-involution with $\mathbb{Z}_{2k}$ is obtained by projecting the quiver matter fields
\begin{equation}\label{orielfieldinv22}
A_i \rightarrow  - i \epsilon_{ij} A_j^*J \qquad B_i \rightarrow  - J B_i^* J \qquad C_i \rightarrow  - i \epsilon_{ij} J C_j^* \qquad D_i \rightarrow  D_i^* \, .
\end{equation}
and it is a theory  with gauge groups $O(2N)\times USp(2N) \times USp(2N) \times O(2N)$.

%\begin{eqnarray}
%& & A_1 = \frac{1}{\sqrt{2}}\Big( \Big( \hbox{Re}(a_1) + \hbox{Im}(a_1) J \Big) + i \Big(\hbox{Re}(a_2)  + \hbox{Im}(a_2) J\Big)\Big)  ,\qquad  B_i = \hbox{Re}(b_i) + \hbox{Im}(b_i) J \nonumber\\
%& & C_1 = \frac{1}{\sqrt{2}}\Big( \Big( \hbox{Re}(c_1) + \hbox{Im}(c_1) J \Big) + i \Big(\hbox{Re}(c_2)  + \hbox{Im}(c_2) J\Big)\Big)  ,\qquad  D_i = \hbox{Re}(d_i) + \hbox{Im}(d_i) J\nonumber
%\end{eqnarray}

%\begin{equation}
%|a_i|^2 - |c_i|^2 = 0 \qquad 2 |a_i|^2 - |b_i|^2 - |d_i|^2 = 0 
%\end{equation}

%\begin{equation}
%a_1 c_2 - c_1 a_2 =0 \qquad b_1 d_2 - b_2 d_1 =0  
%\end{equation}

\subsection{ More general examples}

The construction can be applied to the general ${\cal N}=2$ quiver. In particular, we can orientifold any
quiver by the obvious anti-involution that conjugates all fields. The orientifold action $X_{ab}\rightarrow \Omega_a X_{ab}^* \Omega_b$ is consistent when all $\Omega_a$ are equal and we obtain theories with all orthogonal or all symplectic gauge groups. The one membrane moduli space will be in both cases a Spin(7) manifold of the form $Y/\Theta_k$. This construction can be applied to both chiral and vectorial quivers and typically produces anti-involutions with a locus of fixed points. Actions which fix only the origin  are more difficult to find and require particular choices. We will show  in the next Section how to find other examples by extending the orientifold construction.

\section{Orientifolds with identification of gauge groups}

We can perform more general orientifold projections using $\mathbb{Z}_2$ symmetries of the quiver. The orientifolded
gauge theory is obtained from the parent theory by a certain $\mathbb{Z}_2$ identification of gauge groups and matter fields. This procedure has been utilized systematically in \cite{Franco:2007ii} to construct and classify ${\cal N}=1$ four-dimensional orientifolds theories based on Tilings.   Here we combine the identification of gauge groups and matter fields with an anti-involution in order to obtain an ${\cal N}=1$ theory from a parent ${\cal N}=2$ Chern-Simons theory.
We will not try to be exhaustive and classify all possible models but we restrict to a particular example which exemplifies
the general construction.

\subsection{A Chern-Simons quiver based on  $L^{2,2,2}$ }

\begin{figure}[h]
\centering
\includegraphics[scale=0.6]{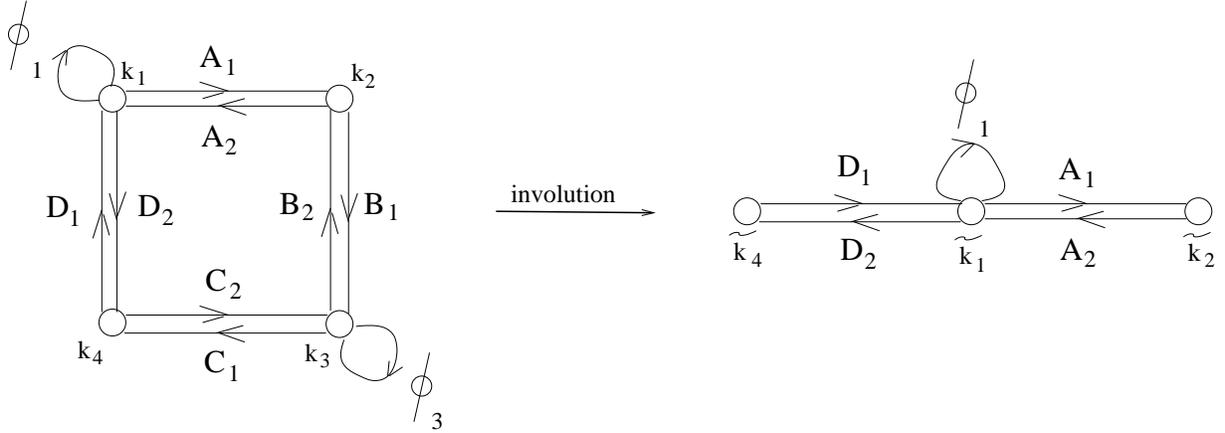}
\caption{Quiver for $\Tilde{L^{2,2,2}}_{(1,-1,1,-1)}$ and its anti-involution.}
\label{L222}
\end{figure}

A particularly simple CY$_4$ is the intersection of two conifold singularities: $x y = w z = t v$. One candidate for the ${\cal N}=2$ theory living on N M2 branes at the tip of this cone is described by the first quiver on the left in Figure \ref{L222}
with superpotential 
\begin{equation} 
W= A_2 A_1 B_1 B_2 - C_2 C_1 D_1 D_2 +\phi_3(C_1 C_2-B_2 B_1) +\phi_1(D_2 D_1- A_1 A_2)
\end{equation}
and Chern-Simons couplings $(1,-1,1,-1)$. The three-fold associated with this quiver and superpotential is part of the $C(L^{a,b,c})$ family \cite{labc}. We refer to the 3d conformal field theory with this specific choice of Chern-Simons as the $\Tilde{L^{2,2,2}}_{(1,-1,1,-1)}$ quiver theory.  To check that  the moduli space of the $\Tilde{L^{2,2,2}}_{(1,-1,1,-1)}$  theory is the intersection of two conifolds let us compute its abelian moduli space.
The abelian F terms imply
\beq \phi_1= B_1B_2=C_1 C_2\, , \qquad\qquad \phi_3 = D_1 D_2 = A_1 A_2 \eeq
The ${\cal N}=2$ abelian moduli space is obtained by modding by the linear combinations of  the (complexified) groups $U(1)_1+U(1)_2$ and $U(1)_1-U(1)_3$. The gauge invariants are
\begin{equation}
x_1= A_1 C_1\, , \,\,\,\,\, x_2 = A_1 A_2\, ,  \,\,\,\,\,  x_3=B_1 D_1\, ,  \,\,\,\,\,  x_4=B_1 B_2\, ,  \,\,\,\,\, 
x_5 = A_2 C_2 \, ,  \,\,\,\,\,  x_6 = B_2 D_2
\end{equation}
and satisfy the two equations
\beq x_2 x_4 = x_1 x_5= x_3 x_6   \eeq
which is exactly the Calabi-Yau fourfold $Y$ we were looking for:  a complete intersection of two quadrics in $\mathbb{C}^6$. For CS couplings $(k,-k,k,-k)$ the moduli space is obtained by modding $Y$ by the remaining discrete group $\mathbb{Z}_k \in U(1)_1-U(1)_2+U(1)_3-U(1)_4$.  

The quiver has a $\mathbb{Z}_2$ symmetry across the diagonal connecting the groups 2 and 4.  The symmetry preserves the choice of Chern-Simons couplings. We can therefore perform an orientifold projection where group 1 is identified with group 3,
the fields $B$ are identified with $A$ and $D$ with $C$. The resulting orientifolded quiver is shown in Figure \ref{L222}.
The gauge groups 2 and 4 are projected with $$A_\mu^{a} =- \Omega_a (A_\mu^a)^T \Omega_a^{-1}\, \qquad a=2,4$$ and
become $O(2N)$ or $USp(2N)$ according to whether $\Omega =I$ or $\Omega=  J$, respectively. The gauge groups 1 and 3 are identified by the orientifold projection $$A_\mu^1 = -\Omega (A_\mu^3)^T \Omega^{-1}$$   and remain 
unitary groups $U(2N)$. The action on fields is
\bea
& A_{1} \rightarrow \hbox{  } \hbox{   }\Omega B_{2}^* \Omega_2^{-1} \, ,\,\,\, A_{2} \rightarrow \hbox{  } \hbox{   }\Omega_2 B_{1}^* \Omega^{-1} \, , \,\,\,  & D_{1} \rightarrow \hbox{  } \hbox{  } \Omega_4 C_{2}^* \Omega^{-1}\, , \,\,\, 
D_{2} \rightarrow \hbox{  } \hbox{  } \Omega C_{1}^* \Omega_4^{-1}  \nonumber\\
& B_{2} \rightarrow\pm  \Omega A_{1}^* \Omega_2^{-1} \, , \,\,\, B_{1} \rightarrow\pm  \Omega_2 A_{2}^* \Omega^{-1}
\, ,\,\,\, & C_{2} \rightarrow \pm  \Omega_4 D_{1}^* \Omega^{-1}\, , \,\,\, C_{1} \rightarrow \pm  \Omega D_{2}^* \Omega_4^{-1}  \nonumber\\
\eea 
where the choice of sign in the second line can be made independently on the pairs $(A,B)$ and the pairs $(C,D)$.
Each choice implies some constraints on the relative choice of $\Omega_{2,4}$ and $\Omega$.

We have three independent models: model I with gauge group $O(2N)_{-2k}\times U(2N)_{2k}\times USp(2N)_{-k}$, model II with  gauge group $O(2N)_{-2k}\times U(2N)_{2k}\times O(2N)_{-2k}$, and model III with  gauge group $USp(2N)_{-k}\times U(2N)_{2k}\times USp(2N)_{-k}$. 
\\

\noindent {\bf Model I:} it  corresponds to an antiinvolution without fixed points.  As usual we analize the case of one membrane, $N=1$. For $N=1$ we have $O(2)\times U(2)\times SU(2)$. The one membrane moduli space can be parameterized by the two-by-two matrices 
\beq
A_{1,2} = \frac{a_{1,2}+b_{2,1}^*}{2} I + \frac{a_{1,2}-b_{2,1}^*}{2 i} J \, ,   D_{1} = \frac{d_{1} + i c_{2}^*}{2} I + \frac{d_{1} -  i c_{2}^*}{2 i} J\, ,   D_{2} = \frac{d_{2} - i c_{1}^*}{2} I + \frac{d_{2} +  i c_{1}^*}{2 i} J
\label{par}
\eeq
As usual, on the moduli space, the D term and F term constraints reduce  to those of the parent ${\cal N}=2$ theory.
We also have four residual abelian groups $SO(2)\subset O(2), SO(2)\subset SU(2), U(1)\times U(1)\subset U(2)$ acting on the moduli space, and  we can arrange for a basis where they act as in the ${\cal N}=2$ theory. One abelian factor acts trivially, two factors are used to mod the moduli space, and the last abelian factor is broken to $\mathbb{Z}_{2k}$
by the Chern-Simons interactions. We have an extra discrete gauge symmetry $i\sigma_3\in SU(2), U(2)$ and $\sigma_3\in O(2)$ acting on the fields as
\bea & a_{1,2}\rightarrow b_{2,1}^* \, , \qquad\qquad & d_{1,2} \rightarrow \hbox{  } \hbox{  } c_{2,1}^* \nonumber\\
& b_{2,1}\rightarrow a_{1,2}^* \, , \qquad\qquad & c_{2,1}\rightarrow - d_{1,2}^* \eea
This transformation is lifted to an anti-involution on $Y$ without fixed points
\bea x_1\rightarrow - x_6^*\, , \qquad  & x_5\rightarrow -x_3^*\, , \qquad  & x_2\rightarrow x_4^*
\nonumber\\ 
 x_6\rightarrow \hbox{  } \hbox{  }x_1^*\, , \qquad  & x_3\rightarrow \hbox{  } \hbox{  } x_5^*\, , \qquad  & x_4\rightarrow x_2^*
\eea
The moduli space is the quotient of $Y/\Theta_k$ by the semi-direct product of the anti-involution with $\mathbb{Z}_{2k}$.
\\

\noindent {\bf Model II and III:} It is easy to see that they both correspond to the anti-involution
\bea x_1\rightarrow x_6^*\, , \qquad  & x_5\rightarrow x_3^*\, , \qquad  & x_2\rightarrow x_4^*
\nonumber\\ 
 x_6\rightarrow x_1^*\, , \qquad  & x_3\rightarrow x_5^*\, , \qquad  & x_4\rightarrow x_2^*
\eea  
which acts on $Y$ with  a fixed locus. 
%The $N=1$ parameterization for model II ($O(2)_{2k}\times U(2)_{2k}\times O(2)_{2k}$)  is obtained from (\ref{par}) by replacing $c_{2,1}\rightarrow i c_{2,1}$, and the parameterization for model III ($SU(2)_{k}\times U(2)_{2k}\times SU(2)_{k}$) by replacing $b_{2,1}\rightarrow i b_{2,1}$. 

\section{Conclusions and Outlooks}

In this paper we presented a large class of ${\cal N}=1$ (2+1) dimensional field theories that are supposed to live on the world volume of N M2 branes at the tip of Spin(7) cones.
They are supersymmetric non-holomorphic field theories with gravity duals. Our procedure is based on an antiholomorphic involution on the  geometrical side  and on orientifold projection on the field theory side. It would be interesting to systematically study the possible ${\cal N}=1$ orientifolds of M2 brane theories. In particular it would be interesting
to perform a general analysis of orientifold M2 theories based on Tilings as it was done in \cite{Franco:2007ii}
for D3 branes. Another nice application would be to extend the ideas of toric/Seiberg duality studied in the context of ${\cal N}=2$ (2+1) dimensional field theories to the ${\cal N}=1$ cases \cite{Amariti:2009rb,Davey:2009sr,Franco:2009sp,Giveon:2008zn,Armoni:2009vv}.  
%  in particular in  cases in which the arrows between the gauge groups are more than two or the gauge groups get identified by the orientifold projection. 
In this paper, we  studied real quotients of CY$_4$ and it would be interesting to see if there is a quiver realization of more general Spin(7) manifolds.

%if we could study also non orbifold Spin(7) manifolds. As a byproduct of our analysis we have found that sometimes the same antiholomorphic involution can be associated to two quivers with different gauge groups. This is somehow reminiscent of ``toric'' duality in which different field theory describe the same geometry. I!
% t would be nice to see if we could extend the ideas of toric duaity studied in the context of ${\cal N}=2$ (2+1) dimensional field theories \cite{.........} to the ${\cal N}=1$ cases.   

\section*{Acknowledgements}

D.F. would like to thank Davide Cassani for stimulating his interest in the subject of Spin(7) geometries and ${\cal N}=1$ Chern-Simons matter theories. Moreover he would like to thank Davide Cassani and Giuseppe Policastro for collaborations at the early stage of this project. It is also a pleasure to acknowledge Amihay Hanany, Yang-Hui He, Alessandro Tomasiello and Angel Uranga for nice and useful discussions. D. ~F.~ is supported by CNRS and ENS Paris. A.~Z.~ is supported in part by INFN and MIUR under contract 2007-5ATT78-002. 
\section*{Appendix}
\appendix
\section{$\mathcal{N}=1$ CS theories}\label{AAP}
In this paper we  studied an ${\cal N}=1$ orientifold projection of the ${\cal N}=2$ theory.
% the field theory living on N M2 branes at the tip of a Spin(7) cone that is obtained by taking a particular antiholomorphic involution on the $CY_4$. The theory living on the M2 branes is supposed to be an $\mathcal{N}=1$ Chern-Simons matter theory obtained with some kind of projection operation acting on the original $\mathcal{N}=2$ Chern Simons matter theory. In the main text we explain how to take this projection,
In this appendix we introduce the $\mathcal{N}=1$ $(2+1)$ dimensional formalism and we explain how to pass from the $\mathcal{N}=2$ to the $\mathcal{N}=1$ notations.

The low energy dynamics of M2 branes at Calabi Yau four-fold toric conical singularities is well described by $(2+1)$ dimensional $\mathcal{N}=2$ Chern-Simons  theories with gauge group $\prod_{i=1}^GU_i(N)$ and bifundamental or adjoint matter fields \cite{Martelli:2008si,Hanany:2008cd}. The typical Lagrangian in $\mathcal{N}=2$ superspace notation is
\begin{eqnarray} 
\label{n2lag}
& & \sum_{a} \frac{k_a}{8\pi} S^{\mathcal{N}=2}_{CS_a}(V_a) - \Tr \Big(\int d^{4} \theta 
\sum_{X_{ab}}
 X_{ab}^{\dagger}e^{-V_a} X_{ab}e^{V_b}\Big) + \int d^{2} \theta W(X_{ab}) + \int d^{2} \bar{\theta} \bar W(X_{ab}^*) \nonumber\\
\end{eqnarray}
%with $\sum_{a} k_a=0$, 
where $V_a$ are the vector superfields, $X_{ab}$ are bifundamental chiral superfields and $S^{\mathcal{N}=2}_{CS_a}$ is the $\mathcal{N}=2$ Chern-Simons action for the $a$-th vector superfield. The Chern-Simons couplings satisfy $\sum_{a} k_a=0$.
In the toric case, the superpotential $W(X_{ab})$ satisfies the following condition: 
every chiral superfield appears exactly twice, one time with plus sign and the other time with minus sign.
Every $V_a$ contains a gauge field, a two-component Dirac spinor and two real scalar fields, $\sigma_a$ and $D_a$. Every $X_{ab}$ contains two complex scalars and a two-dimensional Dirac spinor.
% transforming in the fundamental of the gauge group $a$ and in the anti fundamental of the gauge group $b$, and 
The three dimensional $\mathcal{N}=2$ field content is the dimensional reduction of the four dimensional $\mathcal{N}=1$ superfields. The Lagrangian (\ref{n2lag}) can be intuitively understood as the reduction from four to three dimensions of the field theory living on D3 branes probing a three-fold toric Calabi Yau cone. In the reduction we must substitute the kinetic term for the vector superfields with a Chern-Simons interaction.

In $\mathcal{N}=1$ notations, we need two kinds of  superfields: the spinor superfield $\Gamma_{\alpha}$ and the scalar superfield $\Phi$. The first contains a gauge vector field and a Majorana two-component spinor, while the second contains a Majorana two component spinor and two real scalars. 
The generic  $\mathcal{N}=1$ quiver  Chern-Simons  theory Lagrangian  is
\begin{eqnarray} 
\label{n1lag}
& & \sum_{a} \frac{k_a}{8\pi} S^{\mathcal{N}=1}_{CS_a}(\Gamma^{\alpha}_a) - \int d^{2} \theta_1 \sum_{Y_{ab}} 
 \Tr \Big( \big(D^{\alpha}+i\Gamma^{\alpha}_b \big) Y_{ab}^{\dagger} \big( D_{\alpha}-i\Gamma_{\alpha}^a \big) Y_{ab}\Big) +W^{\mathcal{N}=1}(Y_{ab},Y_{ab}^*) \nonumber\\
\end{eqnarray}
where the first term is the $\mathcal{N}=1$ Chern-Simons action, the second  is the kinetic term for the matter fields and the third  is the $\mathcal{N}=1$ superpotential. It is important to note that the $\mathcal{N}=1$ superpotential
is a real function of the scalar fields. The $Y$s are complex scalar $\mathcal{N}=1$ superfields: $Y=\Phi_1 + i \Phi_2$.

To pass from $\mathcal{N}=2$ to $\mathcal{N}=1$ formalism we need to decompose the $\mathcal{N}=2$ superspace in two copies of an $\mathcal{N}=1$ superspace \cite{csN1},
\begin{eqnarray} 
\label{n2n1sup}
\theta_{\alpha}= \theta_{1\alpha}+ i\theta_{2\alpha},\qquad D_{\alpha}= \frac{1}{2} \big( D_{1\alpha}+  i D_{2\alpha} \big),\qquad  \bar{D}_{\alpha}= \frac{1}{2} \big( D_{1\alpha} -  i D_{2\alpha} \big)
\end{eqnarray}
and keep only  the $\theta_1$ component. The $\mathcal{N}=2$ superfields decompose as
\begin{equation} 
V^a\big|_{\theta_2=0}=0 ,\qquad D_{2\alpha}V^a\big|_{\theta_2=0}= \Gamma_{\alpha}^a ,\qquad D^2_{2}V^a\big|_{\theta_2=0}= R^a ,\qquad X_{ab}\big|_{\theta_2=0}=Y_{ab}\, .
\label{n2n1}
\end{equation}
The $\mathcal{N}=2$ vector superfield $V^a$ decomposes into  an $\mathcal{N}=1$ spinor superfield $\Gamma_{\alpha}^a$ and an $\mathcal{N}=1$ real scalar superfield $R^a$, while the $\mathcal{N}=2$ chiral superfield $X_{ab}$ decomposes into  two real scalar  $\mathcal{N}=1$ superfields Re$\big( X_{ab} \big)\big|_{\theta_2=0}$ and Im$\big( X_{ab} \big)\big|_{\theta_2=0}$ which combine into a complex scalar superfield $Y_{ab}$.
The $R^a$ are $\mathcal{N}=1$ auxiliary real scalar superfields that can be eliminated via their equations of motion. 

When we write   the original $\mathcal{N}=2$ Lagrangian (\ref{n2lag}) in $\mathcal{N}=1$ formalism   the $\mathcal{N}=2$ Chern-Simons and the kinetic terms for scalars become the $\mathcal{N}=1$ Chern-Simons and   kinetic terms. The $\mathcal{N}=1$ superpotential gets contributions from the $\mathcal{N}=2$ Chern-Simons action, the  kinetic terms for chiral fields  and the superpotential 
\begin{equation}
\frac{k_a}{8\pi} S^{\mathcal{N}=2}_{CS_a}\,\,\, \rightarrow \,\,\, -  \frac{k_a}{4\pi} \int d^2\theta_1 R^2_a  
\label{CS2}
\end{equation}
\begin{equation}
-\int d^4\theta X^{\dagger}_{ab} e^{-V_a} X_{ab} e^{V_b}\,\,\,  \rightarrow \,\,\,\int d^2\theta_1\big( Y_{ab} Y^{\dagger}_{ab}R_a  - Y^{\dagger}_{ab} Y_{ab} R_b \big)
\label{cin2}
\end{equation}
\begin{equation}
\int d^2\theta W( X_{ab})+ {\rm c.c}\,\,\,\rightarrow \,\,\,\int d^2\theta_1 W( Y_{ab}) +  W( Y_{ab}^*)
\label{sup2}
\end{equation}
Integrating out the $\mathcal{N}=1$ superfield $R_a$ we get new interactions among the $\mathcal{N}=1$ scalar superfields $Y_{ab}$ depending on the quiver and on the specific values of the CS levels,
\begin{eqnarray} 
\label{n2n1sup}
W^{\mathcal{N}=1}(Y_{ab},Y_{ab}^*)&=& W(Y_{ab}) + W(Y_{ab}^*) +  \sum_{a,k_a\ne0} \frac{k_a}{4\pi}  R_a^2 \nonumber \\
 \frac{k_a}{2\pi} R_a &=& \sum_b Y_{ab} Y_{ab}^\dagger - \sum_c Y_{ca}^\dagger Y_{ca}
\end{eqnarray}
supplemented with the constraints
\beq  \sum_b Y_{ab} Y_{ab}^\dagger - \sum_c Y_{ca}^\dagger Y_{ca}= 0 \eeq
from the gauge groups with zero Chern-Simons coupling.
 
In ${\cal N}=1$ notations the vacuum conditions are obtained by minimizing the superpotential 
with respect to all real superfields. These conditions contain both the F term and the D term constraints of the ${\cal N}=2$ notations. That the two ways of finding a vacuum coincide when the theory has ${\cal N}=2$ supersymmetry follows from the identity  
\bea   \sum_{Y_{ab}} \Tr \Big | \partial_{Y_{ab}} W^{\mathcal{N}=1} \Big |^2  &=&  \sum_{Y_{ab}}\Tr \Big | \partial_{Y_{ab}} W^\dagger  + \sum_b( \sigma_a Y_{ab} -Y_{ab} \sigma_b ) \Big |^2  \nonumber\\
&=& \sum_{Y_{ab}} \Tr | \partial_{Y_{ab}} W |^2 +  \sum_{Y_{ab}} \Tr |\sigma_a Y_{ab} -Y_{ab} \sigma_b|^2
  \eea
valid for all ${\cal N}=2$ quivers. Indeed the first term on the left is the bosonic scalar potential obtained from the $\mathcal{N}=1$ formalism and the last term on the right is the scalar potential coming from the $\mathcal{N}=2$ formalism (\ref{scabos2}).

\end{document}